\documentclass{article}%
\begin{document}
\title{A  Model  with Interacting Composites }
\author{M. Horta\c{c}su$^{*\dag}$ and B.C.L\"{u}tf\"{u}o\={g}lu$^{*}$
\\ \small $^{*}${Department of
Physics, Istanbul Technical University, Istanbul, Turkey}.
 \\ \small$^{\dag}${Feza G\"{u}rsey Institute, Istanbul,
 Turkey.}\\\small$^{\dag}${hortacsu@itu.edu.tr} \\ \small$^{\ddag}${bcan@itu.edu.tr}}
\date{\today}%
\maketitle
\begin{abstract} \noindent We show that we can construct
 a model  in $3+1$ dimensions where
only composite scalars take place in physical processes as
incoming and outgoing particles, whereas constituent spinors only
act as intermediary particles. Hence while the spinor-spinor
scattering goes to zero, the scattering of composites gives
nontrivial results. \vspace{5mm}
\end{abstract}
\section{Introduction}
Fermions are an essential ingredient in nature. It is an ever
repeating idea to build a model of nature using only fermions,
where all the observed bosons are constructed as composites of
these entities. In solid state physics electrons, fermions in
character, come together to form bosons \cite{ref1}. Heisenberg
spent years to formulate a "theory of everything" using only
fermions \cite{ref2}. Another attempt in this direction came with
the work of G\" ursey, \cite{ref3}, where a non-polynomial
Lagrangian was written to describe self-interacting fermions.
Kortel found solutions to this theory \cite{ref4} which were shown
to be instantonic and meronic solutions much later \cite{ref5}.

One of us, with collaborators, tried to make quantum sense of this
model a while ago \cite{ref6- ref8}, finding that even if these
attempts are justified, this model went to
 a trivial model as the cut-off is removed. We calculated
\cite{ref9} several processes involving incoming and outgoing
spinors which gave exactly the naive quark model results, missing
the observed logarithmic behaviour predicted by  QCD calculations.

During the last twenty years, many papers were written on making
sense out of "trivial models", interpreting them as effective
theories without taking the cutoff to infinity.  One of these
models is the Nambu-Jona-Lasinio model \cite{ref10}.  Although
this model  is shown to be a trivial  in four dimensions
\cite{ref11-ref13}, since the coupling constant goes to zero with
a negative power of the logarithm of the ultraviolet  cut-off, as
an effective model in  low energies it gives us important insight
to several processes. In QCD, the studies of hadron mass
generation through spontaneous symmetry breaking, important clues
to results of the nuclear pairing interaction and the approximate
validity of the interacting boson model can be cited as some
examples.

There were also attempts to couple the Nambu-Jona-Lasinio  model
to a gauge field, the so called gauged Nambu-Jona-Lasinio model
\cite{ref14,ref15}  to be able  to get a non-trivial field theory.
It was shown that if one has sufficient number of fermion flavors,
such a construction is indeed possible \cite{ref16}.  Actually the
Nambu-Jona-Lasinio model was constructed based on an  analogy with
the BCS theory of superconductivity, where fermions come together
to form the bosonic interaction necessary to explain the physical
phenomena. \cite{ref17}

Here we want to give a new interpretation of our old work. First
we see that the polynomial form of the original model really does
not correspond to it in the exact sense.  The two versions have
obey different symmetries. Then  we  go to higher orders  in our
calculation in the new version, beyond the one loop for the
scattering processes. It is shown that by using the
Dyson-Schwinger and Bethe-Salpeter equations some of the
fundamental processes can be better understood. We see that while
the non-trivial scattering of the fundamental fields is not
allowed, bound states can scatter from each other with non-trivial
amplitudes. This phenomena is another example of treating the
bound states, instead of the principal fields, as physical
entities, that go through physical processes such as scattering.

In our model we  need an infinite renormalization in one of the
diagrams. Further renormalization is necessary at each higher
loop, like any other renormalizable model. The difference between
our model and other renormalizable  models lies in the fact that,
although our model is a renormalizable one  using naive
dimensional counting arguments, we have only one set of diagrams
which is divergent. We need to renormalize only one of the
coupling constants by an infinite amount. This set of diagrams,
corresponding to the scattering of two bound states to two bound
states, have the same type of divergence, i.e.
${{1}\over{\epsilon}}$ in the dimensional regularization scheme
for all odd number of loops. The contributions from even number of
diagrams are finite, hence require no infinite renormalization.
The scattering of two scalars to four, or to any higher even
number of  scalars is finite, as expected to have a renormalizable
model, whereas production of spinors from the scattering of
scalars go to zero as the cut-off is removed.

We will outline the model as is given in Refs. \cite{ref6} and
\cite{ref7} in Section I and give our new results in subsequent
sections .

\section{The Model}

We start with the Lagrangian of the model given as
\begin{equation}
L = {i\overline{\psi}} \partial \!\!\!/ \psi + g {\overline{\psi}}
\psi \phi +\lambda ( g{\overline {\psi}} \psi -a\phi^{3} ).
\end{equation}
Here the only terms with kinetic part are the spinors.
Here $\lambda$ is a Lagrange multiplier field, $\phi $ is a
scalar field with no kinetic part, $g$ and $a$ are coupling constants. This expression
contains two constraint equations, obtained from writing the Euler
-Lagrange equations for the $\lambda$ and $\phi$ fields.
\begin{equation}
 g{\overline{\psi}} \psi -3 \lambda a\phi^{2} =0,
\end{equation}
 and
\begin{equation}
 g{\overline{\psi}} \psi -a\phi^{3} =0 .
\end{equation}
The Lagrangian given above is just an attempt in writing the original G\" ursey
Lagrangian
\begin{equation}
 L={i\overline{\psi}} \partial \!\!\!/ \psi + g' ({\overline{\psi}} \psi)^{4/3} ,
\end{equation}
in a polynomial form.

We see that the $\gamma^{5} $ invariance of the original
Lagrangian is retained in the form written  in Eq. $(1)$. In this
form, when $\psi$ is sent to $\gamma^{5} \psi$, the scalar fields
$\phi$ and $\lambda$ are sent to their negatives ( minus times the
field).  This discrete symmetry prevents $\psi$ from acquiring a
finite mass in higher orders.

 We see that these two models are not equivalent since the latter does not obey one symmetry obeyed by the former one.  If we take the original Lagrangian
\begin{equation}
L = {\bar \psi_i \partial \llap{$/$} \psi}+(\bar \psi
\psi)^{4/3}+s(\bar \omega (\bar \psi \psi+\phi^{3}))~,
\end{equation}
and define a symmetry operation $s$ where
  $s\bar \omega = \lambda, s\lambda = 0, s\phi = \omega, s\omega = 0, s\psi = s\bar \psi = 0$,
  so that $L$ is invariant under $s$ . If we replace the original Lagrangian by that given in equation(1), by
 replacing $(\bar \psi  \psi)^{4/3}$ by a combination of $\phi ^{4}, \phi \bar \psi \psi$ we see that this symmetry is not retained.  We, therefore, take the second model as a model which only approximates the original one, without claiming equivalence.  It is  a constrained model which will replace in the original model only in a "naive" sense

To quantize the latter system consistently we proceed through the
path integral method.  In addition to the usual spinor-Dirac
primary constraints, fixing the momenta corresponding to the
spinor fields $\psi$ and $\overline {\psi}$, we have two new
primary constraints, setting the canonical momenta corresponding
to the scalar fields $\lambda$ and $\phi$ equal to zero. The
primary Hamiltonian is obtained by  adding these four constraints
multiplied by arbitrary constants  to the canonical Hamiltonian,
obtained  from the Lagrangian given in Eq. $(1)$.  The consistency
requirement of all the primary constraints, which is setting the
Poisson bracket of the constraint equations with the primary
hamiltonian equal to zero, results in two new, secondary
constraints, given by our Eqs. $(2)$ and $(3)$ . When we calculate
the Poisson bracket of these constraints with the primary
Hamiltonian to check whether additional constraints are present,
we see that the system is closed, determining all the arbitrary
constants in the primary Hamiltonian.

Next we compute the determinant of the Poisson brackets of all the
second class constraints, the so called Faddeev-Popov determinant.
We see that the spinor-Dirac constraints, resulting from the
canonical momenta of the spinor fields has no field dependent
contribution to the   Faddeev-Popov determinant. This determinant
is given as
\begin{equation}
\Delta_{F} = [ det \lbrace \theta_{i} , \theta_{j} \rbrace_{P}
]^{1/2} = det \phi^{4} .
\end{equation}
the field dependent contribution coming from the constraints in
Eqs.$(2)$ and $(3)$.

We write the generating functional for the Green's functions of
the model as
\begin{equation}
 Z = \int D \pi D\chi \delta ( \theta_{i}) \Delta_F
exp {(-i\int ( \dot{\chi} \pi -H_c))}.
\end{equation}
Here $\chi$ is the generic symbol for all the fields , $\pi$ is
the generic symbol for all momenta and $\theta $ is the generic
symbol for all the constraints in the model. Performing all the
momenta integrals we obtain
\begin{equation}
 Z = \int D \overline{\psi} D \psi D \phi D \lambda \left(
{{\Delta_F}\over{3 det \phi^2}} \right)exp ( i \int L' d^4 x) ,
\end{equation}
 where
\begin{equation}
  {{\Delta_F}\over{det \phi^2}} = det \phi^2 ,
\end{equation}
 This  contribution is inserted into the Lagrangian using ghost fields $c$ and $c^*$,
 and the resulting lagrangian reads
 \begin{equation}
L'' = {\overline{\psi}}[\partial \!\!\!/  + g  (\phi +\lambda ) ]
\psi - a\lambda \phi^{3}  + ic^*\phi^{2}
c.
\end{equation}
We can rewrite  this expression by defining
\begin{equation}
\Phi = \phi+\lambda,
\end{equation}
\begin{equation}
\Lambda =\phi-\lambda,
\end{equation}
 as
\begin{equation}
 L'' = {i\overline{\psi}}[ \partial \!\!\!/  + g  \Phi]
\psi -{{a }\over{16}} (\Phi^{4}+2\Phi^{3}\Lambda-2\Phi\Lambda^{3}-\Lambda^{4})+{i\over{4}}c^*(\Phi^{2}+2\Phi\Lambda +\Lambda^{2}) c.
\end{equation}
By this transformation the $\Lambda$\begin{scriptsize}\end{scriptsize} field is decoupled
from the spinor sector of the lagrangian.

The integration over the spinor fields in the functional yields
the effective action which is expressed in terms of $\Phi,
\Lambda$ and $c $, $c^*$ fields only.
\begin{eqnarray}
 S_{eff} = - Tr ln ( i \partial \!\!\!/ + g \Phi )
 + \int d^4 x \left[{{a }\over{16}} (\Phi^{4}+2\Phi^{3}\Lambda-2\Phi\Lambda^{3}-\Lambda^{4})\right. \nonumber \\
 -\left.{i\over{4}}c^*(\Phi^{2}+2\Phi\Lambda +\Lambda^{2}) c\right] .
\end{eqnarray}
The condition to get rid of  the tadpole contribution  , which is
setting the first derivative of the effective action with respect
to the $\Phi$ and $\Lambda$ fields to zero, gives us two equations
\begin{equation}
 {{-ig}\over{(2\pi)^4}} Tr \int {{d^4p}\over{ p\!\!\!/ -gv}}-{{a }\over{8}}(2v^{3}+3v^{2}s-s^{3})=0,
\end{equation}
 and
\begin{equation}
 {{a }\over{8}}(v^{3}-3vs^{2}-2s^{3})= 0 .
\end{equation}
In these expressions -v and s are the vacuum expectation values of
the fields $\Phi$ and $\Lambda$ respectively and the vacuum
expectation value of the ghost fields are set to zero.

A consistent solution satisfying both equations is
\begin{equation}
 s=v=0,
\end{equation}
Since the $\gamma^5$ symmetry is not dynamically broken, no mass
is generated for the fermion dynamically.  In this respect this
model differs from the famous Gross-Neveu model, \cite{ref18}
where this dynamical breaking takes place.  It also differs from
the Nambu-Jona-Lasinio model.  The main reason for this behaviour
is the conformal invariance present in the  model . G\" ursey's
original intention was to construct a conformal invariant model,
at least classically. We find that upon quantization of our
approximate model at least one phase exists which respects this
symmetry .

The fermion propagator is the usual Dirac propagator in lowest
order, as can be seen from the Lagrangian.  The second derivative
of the effective action with respect to the $\Phi$ field gives us
the induced inverse propagator for the $\Phi$ field , with the infinite part given as
\begin{equation}
Inf \left[ {{ig^2}\over{ (2\pi)^4}} \int {{d^4 p}\over {p\!\!\!/(p\!\!\!/+q\!\!\!/)}}\right]=
 {{g^2  q^2}\over {4\pi \epsilon}}
\end{equation}
Here dimensional regularization is used for the momentum integral
and $\epsilon = 4-n$.  We see that the $\Phi$ field propagates as
a massless field.

When we study the propagators for the other fields, we see that
no linear or quadratic term in $\Lambda$ exists, so the one loop
contribution to the $\Lambda$ propagator is absent. Similarly the
mixed derivatives of the effective action with respect to
$\Lambda$ and $\Phi$  is zero at one loop, so no mixing between
these two fields occurs. We can also set the propagators of the
ghost fields to zero, since they give no contribution in the one
loop approximation.  The higher loop contributions are absent  for
these fields.

\section{Dressed Fermion Propagator}

In this section we calculate the above results in higher orders.
To justify our result that no mass is generated for the fermion
we may study the Bethe-Salpeter equation obeyed for this
propagator. The Dyson-Schwinger equation for the spinor propagator
is written as
\begin{equation}
 iA p\!\!\!/ +B = i p\!\!\!/ + 4\pi \epsilon \int {{d^4 q}\over {( i A q\!\!\!/ +B)(p-q)^2}} .
 \end{equation}
Here $iA p\!\!\!/ +B$ is the dressed fermion propagator. We use
the one loop result for the scalar propagator. After rationalizing
the denominator, we can take the trace of this expression over the
$\gamma$ matrices to give us
\begin{equation}
 B= 4\pi \epsilon \int  d^4 q {{B}\over {(A^2 q^2 +B^2)(p-q)^2}} .
\end{equation}
 The angular integral on the right hand side can be performed to
give
\begin{equation}
B= 4\pi \epsilon \left[ \int_{0}^{p^2} dq^2 {{q^2 B}\over{ p^2(A^2
q^2+ B^2)}}+ \int_{p^2}^{\infty} dq^2  {{ B}\over{ (A^2 q^2+
B^2)}}\right].
\end{equation}
 If we differentiate this expression with respect
to $p^2$ on both sides, we get
\begin{equation}
{{dB}\over{dp^2}}= -4\pi \epsilon \left[ \int_{0}^{p^2} dq^2 {{q^2
B}\over{ (p^2)^2(A^2 q^2+ B^2)}}\right] .
\end{equation}
 This integral is clearly finite.  We get zero for the right hand side as $\epsilon$
goes to zero. Since mass is equal to zero in the free case we get
this constant equal to zero. This choice satisfies the Eq. $(19)$.

The similar argument can be used to show that $A$ is the dressed
spinor propagator is a constant. We multiply Eq. $(18)$ by
$p\!\!\!/$ and then take the trace over the spinor indices. We end
up with
\begin{equation}
p^2 A = p^2 + 4\pi \epsilon \left[ \int_{0}^{p^2} dq^2 \left(
{{(q^4) A}\over{p^2 (A^2 q^2+ B^2)}}+ \int_{p^2}^{\infty} dq^2
{{p^2 A}\over{ (A^2 q^2+ B^2)}}\right)\right].
\end{equation}
We divide both sides by $p^2$ and differentiate with respect to
$p^2$.  The end result
\begin{equation}
{{dA}\over{dp^2}} = - 8\pi \epsilon  \int_{0}^{p^2} dq^2 \left(
{{(q^4) A}\over{ (p^2)^3(A^2 q^2+ B^2)}}\right)
\end{equation}
shows that $A$ is a constant as $\epsilon$ goes to zero.  Since
the integral is finite, it equals unity for the free case, we take
$A=1$.

\section{Higher Orders}

If our fermion field had a color index $i$ where $i=1...N$, we
could perform an 1/N expansion to justify the use of only ladder
diagrams for higher orders for the scattering processes. Although
in our model the spinor has only one color, we still consider only
ladder diagrams anticipating that one can construct a variation of
the model with N colors.

 We first see that we do not need infinite
regularization for the $< {\overline {\psi}} \psi \phi> $ vertex.
The one loop correction to the tree vertex involves two fermion
and one $\phi$ propagator and one integration.  The infinity
coming from the momentum integration is cancelled by the
$\epsilon$ in the $\phi$ propagator. All the higher order
contributions vanish because the powers of $ \epsilon$ exceed the
number of infinities coming momentum integrations. Indeed there is
only one  momentum integration that results in an infinity. We see
that there is only a finite renormalization of the spinor-scalar
coupling constant $g$.

We come to the same result after we write the Dyson-Schwinger
equation for this vertex. We need the result of the four fermion
scattering kernel to be able to perform this calculation. There is
no four fermion coupling in our Lagrangian; so, this process will
use at least one scalar propagator.  Since the scalar particle
propagator has a $\epsilon$ factor, this process vanishes as
$\epsilon$ goes to zero. All higher orders, including the one loop
contribution also vanish as $\epsilon$ goes to zero, since they
have higher powers of $\epsilon$.

We can justify our claims also by writing the Bethe-Salpeter
equations for this process. The Bethe-Salpeter equation for the
fermion interaction reads as

\begin{eqnarray}
G^{(2)}(p,q;P) = G^{(2)}_{0} (p,q;P)\hspace{6cm} \nonumber \\+
{{1}\over{(2\pi)^8 }} \int d^4 p' d^4 q' G^{(2)}_{0} (p,p';P)
K(p',q';P) G^{(2)} (q',q;P).
\end{eqnarray}
Here $G^{(2)}_{0} (p,q;P)$ is two non-crossing spinor lines, $
G^{(2)} (p,q;P) $ is the proper four point function. $K$ is the
Bethe-Salpeter kernel.

We note that this expression involves the four spinor  kernel
which we found to be of order $\epsilon $.  This expression can be
written in the quenched ladder approximation \cite{ref1}, where
the kernel is seperated into a scalar propagator with two spinor
legs joining the proper kernel.  If the proper kernel is of order
$\epsilon$ , the loop involving two spinors and a scalar
propagator can be at most finite that makes the whole diagram in
first order in $\epsilon$.    This fact also shows that there is
no nontrivial spinor-spinor scattering.

We use this result in calculating the Dyson-Schwinger equation for
the spinor-scalar vertex. This vertex involves the finite coupling
constant $g$ plus the diagram where the scalar particle goes into
two spinors which go to the four spinor Kernel. Here the
$\epsilon$ factor coming from the Kernel is cancelled with the
loop integration. The loop does not involve any scalar
propagators, so it diverges as ${{1}\over {\epsilon}}$. The result
is finite renormalization of the three point vertex. Hence the
spinor-scalar coupling constant does not {\it run}.

We see that the only infinite remormalization is needed for the
four $\phi$ vertex; hence the coupling constant for this process
{\it run}. The first correction to the tree diagram is the box
diagram. This diagram has four spinor propagators and give rise to
a ${{1}\over {\epsilon}} $ type divergence. Since we included the
four $\phi$ term in  our original lagrangian, we can renormalize
the coupling constant of this vertex to incorporate this
divergence. The finite part of this diagram is just a constant,
renormalizing the initial coupling constant by a finite amount.
There are no higher infinities for this vertex. The two loop
diagram contains a $\phi$ propagator which makes this diagram
finite. The three-loop diagram is made out of eight spinor and two
scalar lines. At worst we end up with a first order infinity of
the form ${{1}\over{\epsilon}}$ using the dimensional
regularization scheme.  Higher order ladder diagrams give at worst
the same type of divergence. This divergence for the four scalar
vertex  can be renormalized using standart means.

\section{Conclusion}

As a result of this analysis we end up with a model where there is
no scattering of the fundamental fields, i.e. the spinors, whereas
the composite fields, the scalar field, can take part in a
scattering process. Here we do not give the exact expression for
this amplitude, but it will be a series in $a$ and even powers of
$g$, starting with $g^4$.

We can also have scattering processes where two scalar particles
go to an even number of scalar particles.  In the one loop
approximation all these diagrams give finite results, like the
case in the standard Yukawa coupling model. Since going to an odd
number of scalars is forbidden by the $\gamma^5$ invariance of the
theory, we can also argue that scalar $\phi$ particles can go to
an even number of scalar particles only. This assertion is easily
checked by diagrammatic analysis.

Any diagram  which describes the process of producing spinor
particles out of two scalars contains scalar propagators.  The
lowest of these diagrams where two scalars go to two spinors
vanish
 since it either involves a triangle diagram
made out of spinors, or  a box diagram, made out of three spinors
and one scalar. It vanishes due to fall of the scalar propagator
in the latter case, although it is not zero unless the cut-off is
removed . The diagram which involves the production of four
spinors out of two scalars carries two scalar propagators and the
diagram vanishes with the first power of $\epsilon$.

As a result of our calculations we find a model which is trivial
for the constituent spinor fields, whereas finite results are
obtained for thee scattering of the composite scalar particles.
The coupling constant for the scattering of the composite
particles run, whereas the coupling constant for the spinor-scalar
interaction does not run.

In the classical model, described by the Lagrangian given by Eq.
$(4)$, we used one coupling constant $g'$, which is divided into
two as $g$ and $a$ in Eq. $(1)$.  We see that these two behave
differently in the quantum case.

Our model is a toy model. We could not yet find a physical system that
is effectively described by it.

{\bf{Acknowledgement}}: We thank Ferhat Ta\c sk\i n, Kayhan \"{U}lker, and an unknown correspondence for discussions
and both scientific and technical assistance throughout this work.
The work of M.H. is also supported by TUBA, the Academy of
Sciences of Turkey. This work is also supported by TUBITAK, the
Scientific and Technological Council of Turkey.

\end{document}